\documentstyle[aps,prl,twocolumn,epsf]{revtex}
\input epsf
\begin{document}
\noindent
{\bf Kolomeisky, Newman, Straley and Qi Reply:} \ Bhaduri and Sen (BS) 
\cite{bs} criticize our theory of low-dimensional inhomogeneous Bose liquids 
\cite{KNSQ} on two points which we refute below.

1.  BS object to the way our energy functional depends on the
gradient of the density, and propose two alternatives.  
As they admit, the first of these (no dependence) fails to
describe what happens at a hard wall.  It is also inferior to
our theory in representing what happens near a classical turning
point (see Figure 1 of our paper\cite{KNSQ}), and is unable to
describe the important quantum mechanical effect of particle 
tunnelling through a classical barrier, which is present in our theory.
 
Their second choice is even worse: the term $- (\rho ^ {\prime})^2 /\rho$
means that gradients lower the energy: the ground state is unstable 
against spontaneous waviness.  The corresponding nonlinear Schr\"odinger 
equation does not have physically sensible solutions.

Figure 1 compares the behavior near $x = 0$ of the
exact density function for a dark
soliton\cite{GW}
$n(x) = (2N + 1)/2L - \sin[2\pi(2N + 1)x/L]/[2L\sin (2\pi x/L)]$ 
with the approximate expression
$
n_{approx}(x) = (2N/L)[3\coth^{2}(k_{F}x) - 1]^{-1}
$
which is Eq. (12) of our Letter, with $\beta = 0$
(however, please note that $k_{F} = 2 \pi N/L$ for
this example, because we only fill every other
single particle state).

\begin{figure}[htbp]
\epsfxsize=2.5in
\vspace*{-0.1cm}
\epsfbox{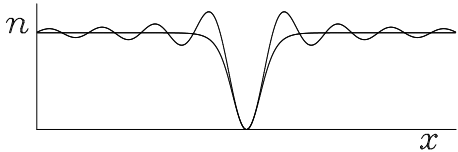}
\vspace*{0.1cm}
\caption{Exact and approximate density profiles near a soliton.
The exact solution has additional short wavelength
structure not captured by the long-wavelength theory.}
\end{figure} 

The alternate forms for the energy functional proposed
by BS cannot describe this stationary soliton at all.
The density gradient
terms in our theory are appropriate and necessary.

Gray soliton solutions of the system of free fermions have been 
generated numerically \cite {GW} although their exact analytical form is 
unknown.  Here our theory is the only source of analytical results.  

Another example which verifies the starting point of our theory and can be
compared with the exact results concerns self-similar solutions whose 
existence was pointed out in our Letter \cite{KNSQ}. Consider a dilute 
system of particles placed in a harmonic trap of dimensionless frequency
$w$. In the dimensionless variables of our Letter it is described by

\begin{equation}
\label{harmeom}
2i\partial _{\tau}f = -\partial _{y}^{2}f + [|f|^{4}
+ w^{2}y^{2}]f \ 
\end{equation}
The self-similar solution $f=Ae^{i\theta}$ derived from (\ref{harmeom}) has 
the form

\begin{equation}
\label{sss}
A=\rho(\tau)^{-1/2}h(y/\rho(\tau)) \ , \ \ \ \theta = \theta _{0}(\tau)
+ {1\over 2} {d\ln \rho \over d\tau} y^{2} \ ,
\end{equation} 
where the functions $\rho (\tau)$ and $h(v)$ obey the equations
\begin{eqnarray}
\label{auxeqsa}
{d^{2}\rho \over d\tau ^{2}} & = & -w^{2}\rho + {\gamma \over \rho ^{3}} \\ 
\label{auxeqsb}
{d^{2}h \over dv^{2}} & = & -\delta h + h^{5} + \gamma v^{2}h 
\end{eqnarray}
where $\gamma$ and $\delta$ are arbitrary constants. Eq.(\ref{auxeqsb}) for 
the scaling function $h(v)$ has localized solutions only for $\delta > 0$
and $\gamma \ge 0$: for $\gamma = 0$ an explicit analytical solution can be
written down, while for $\gamma > 0$, (\ref{auxeqsb}) has the same 
functional form as the equation we encountered in determining
the density profile in the harmonic trap [cf. Eq. (6) of the Letter
for $V=m\omega^{2}x^{2}/2$].

The dynamics of the length scale $\rho (\tau)$ can be understood by viewing
(\ref{auxeqsa}) as a fictitious classical mechanics problem in the 
potential $U=w^{2}\rho ^{2}/2 + \gamma/2\rho^{2}$. This analogy implies
that an initially localized cloud of particles in free space
($w=0$) expands asymptotically in a ballistic fashion: 
$\rho (\tau) \sim \tau$.  These conclusions are in complete agreement with 
the results of a full quantum-mechanical calculation \cite{td} of expanding
cloud of free fermions.   In the presence of the confining potential 
($w \ne 0$) the scale $\rho (\tau)$ oscillates between maximum and minimum 
values: for $\gamma=0$ the dynamics of $\rho $ is the same as for a harmonic
oscillator of frequency $w$.

2.  BS state that current experimental systems are not dilute.   
However, appropriate experiments will be possible in the future
(their feasibility has been discussed\cite{hat}) to which our results are
relevant.  Our approach
can provide a useful insight even where it is not quantitatively
correct.

This work was supported by the Thomas F. Jeffress and Kate Miller Jeffress
Memorial Trust.

\bigskip
{\obeylines
\noindent Eugene B. Kolomeisky, T. J. Newman, J. P. Straley, and Xiaoya Qi
Department of Physics
University of Virginia 
382 McCormick Road
P. O. Box 400714 
Charlottesville, VA 22904-4714, USA
}

\medskip

\noindent
PACS numbers: 03.75.Fi, 05.30.Jp, 32.80.Pj

\end{document}